\documentclass{jps-cp}
\usepackage{txfonts} %Please comment out this line unless the txfonts package is availabe in your LaTeX system.
\usepackage{bm}
\allowdisplaybreaks

\title{Dynamically Integrated Transport Approach for High-Energy Nuclear Collisions at High Baryon Density}

\author{
  Koichi \textsc{Murase},$^{1}$
  Yukinao \textsc{Akamatsu},$^{2}$
  Masayuki \textsc{Asakawa},$^{2}$
  Tetsufumi \textsc{Hirano},$^{1}$
  Masakiyo \textsc{Kitazawa},$^{2,3}$
  Kenji \textsc{Morita},$^{4,5}$
  Yasushi \textsc{Nara},$^{6}$
  Chiho \textsc{Nonaka}$^{7,8}$
  and Akira \textsc{Ohnishi}$^{9}$}

\inst{
  $^{1}$Department of Physics, Sophia University, Tokyo 102-8554, Japan\\
  $^{2}$Department of Physics, Osaka University, Toyonaka, Osaka 560-0043, Japan\\
  $^{3}$J-PARC Branch, KEK Theory Center, Institute of Particle and Nuclear Studies,
    KEK, 203-1, Shirakata, Tokai, Ibaraki, 319-1106, Japan\\
  $^{4}$Institute of Theoretical Physics, University of Wroclaw, 50204 Wrocław, Poland\\
  $^{5}$iTHES Research Group, RIKEN, Saitama 351-0198, Japan\\
  $^{6}$Akita International University, Yuwa, Akita-city 010-1292, Japan\\
  $^{7}$Department of Physics, Nagoya University, Nagoya 464-8602, Japan\\
  $^{8}$Kobayashi Maskawa Institute, Nagoya University, Nagoya 464-8602, Japan\\
  $^{9}$Yukawa Institute for Theoretical Physics, Kyoto University, Kyoto 606-8502, Japan}

\email{
  murase@sophia.ac.jp,
  akamatsu@kern.phys.sci.osaka-u.ac.jp,
  yuki@phys.sci.osaka-u.ac.jp,
  hirano@sophia.ac.jp,
  kitazawa@phys.sci.osaka-u.ac.jp,
  kmorita@yukawa.kyoto-u.ac.jp,
  nara@aiu.ac.jp,
  nonaka@hken.phys.nagoya-u.ac.jp,
  ohnishi@yukawa.kyoto-u.ac.jp}

\recdate{January 18, 2019}

\abst{
  To explore the structure of the QCD phase diagram
  in high baryon density domain,
  several high-energy nuclear collision experiments
  in a wide range of beam energies
  are currently performed or planned
  using many accelerator facilities.
  In these experiments
  search for a first-order phase transition and the QCD critical point
  is one of the most important topics.
  To find the signature of the phase transition,
  experimental data should be compared to
  appropriate dynamical models which quantitatively describe
  the process of the collisions.
  In this study we develop a new dynamical model
  on the basis of the non-equilibrium hadronic transport model JAM
  and 3+1D hydrodynamics.
  We show that the new model reproduce well
  the experimental beam-energy dependence of hadron yields and particle ratio
  by the partial thermalization of the system
  in our core--corona approach.}

\kword{high-energy nuclear collisions, beam energy scan, dynamical initialization, core--corona separation}
\begin{document}
\maketitle

\section{Introduction}

One of the important goals in quantum chromodynamics (QCD) is to reveal the structure of the phase diagram.
The phase diagram is drawn in a plane with the horizontal and vertical axis
being the temperature $T$ and the baryon chemical potential $\mu_B$, respectively.
With the vanishing baryon chemical potential,
the transition from a hadron gas to quark gluon plasma (QGP)
is known to be crossover by lattice QCD calculations.
For the finite baryon chemical potential,
the phase structure is less known
because such first-principles calculations are not available due to the sign problem.
To explore the high baryon density domain of the phase diagram
by high-energy nuclear collision experiments,
Beam Energy Scan (BES) programs
at the Relativistic Heavy Ion Collider (RHIC)
at BNL %at Brookhaven National Laboratory (BNL)
and NA61/SHINE experiment at the Super Proton Synchrotron (SPS) at CERN
are ongoing.
Several future experiments
such the CBM experiment at FAIR, MPD at NICA, CEE at HIAF,
and a heavy-ion program at J-PARC (J-PARC-HI)
are also currently planned to search a wider range of the beam energy
to explore a broader region of the phase diagram.
The most interesting topic in these experiments
is search of the first-order phase transition and the QCD critical point
which are predicted by some theoretical models~\cite{ref:critical-point}.

In experiments, the direct observable
is just momentum distributions of final state hadrons.
To reconstruct from the hadron spectra
the information of the high baryon density matter
created in the middle stage of collision reactions,
we need an appropriate dynamical model to quantitatively
describe the whole process of collision reactions
in a wide range of the beam energy.
%% We emphasize that in the lower-energy collisions compared to the RHIC top energy,
%% the created matter is highly inhomogeneous so that
%% merely a part of the system is thermalized to form a high baryon density matter
%% yet the other part is still in a non-equilibrium dilute hadron gas.
%% To properly describe this inhomogeneous dynamics,
We present a new method to dynamically integrate
a hadronic transport model and a hydrodynamic model,
which have been used in describing lower-energy and higher-energy collisions, respectively.

\section{Model}
In our new model, JAM+Hydro,
we dynamically integrate the hadronic transport model JAM~\cite{ref:JAM} and ideal hydrodynamics~\cite{ref:PAPER}.
%% Here we explain three aspects of our dynamical integration,
%% {\it i.e.},
%% dynamical intiailization,
%% dynamical core--corona separation
%% and dynamical coupling of two models.
In the lower-energy collisions for the high baryon density region,
the colliding nuclei take finite time to pass through each other,
which requires a dynamical description of the initial dynamics.
%% the Lorentz contraction of the colliding nuclei is insufficient so that
%% the initial dynamics before the nuclei pass through each other
%% needs dynamical descripation.
Here we adopt the idea of the {\it dynamical initialization}~\cite{ref:dynamical-init} in which
hydrodynamics initially starts with vacuum,
and then the fluids are created by source terms described by the transport model.
Also, in the lower-energy collisions,
%% the fraction of the high-density part of the system which gets thermalized
the fraction of the thermalized part in the system
%% becomes smaller compared to that in the higher-energy collisions.
becomes smaller.
Therefore, we need to separate the thermalized part ({\it core})
and the other non-equilibrium part ({\it corona})~\cite{ref:core-corona-sep}
to describe each part with an appropriate model,
{\it i.e.}, hydrodynamics for core and a non-equilibrium transport model for corona.
In our model we make such separation in both of space and time dynamically,
which we call {\it dynamical core--corona separation}.
Finally the interaction between the thermalized part and the non-equilibrium part
is also important.
%% To introduce the interaction between hydrodynamics and the transport model,
%% we need to simultaneously solve the time evolutions of the two models
%% which are dynamcially coupled to each other,
%% for which we use the term {\it dynamical coupling} of the two models.
Here we need to consider the {\it dynamical coupling} of the two models
whose time evolution is simultaneously solved.

%% Embracing these ideas in our minds, we construct our model
%% by combining JAM and ideal hydrodynamics.
The corona part of the system is described by JAM cascade,
a microscopic transport model
which contains the binary collisions of particles,
resonance decays and 
string formation and fragmentation.
The core part is described by ideal hydrodynamics
with a phenomenological equation of state with a first-order phase transition, EOS-Q~\cite{ref:EOSQ}.
The core--corona separation is controlled by two parameters,
%% the fluidization energy density $e_\mathrm{f}$
%% and the particlization energy density $e_\mathrm{p}$,
the fluidization and particlization energy density $e_\mathrm{f}$ and $e_\mathrm{p}$,
for which we use $e_\mathrm{f}=e_\mathrm{p}=0.5\ \text{GeV}/\text{fm}^3$
unless otherwise specified.
%% Hydrodynamics is initialized and interact with JAM cascade through the source terms:
Hydrodynamics is initialized through the source terms:
\begin{align}
  \partial_\mu T^{\mu\nu}_f &= J^\nu, &
  \partial_\mu N_f^\mu &= \rho,
\end{align}
where $T^{\mu\nu}_f$ and $N_f^\mu$ are
the energy--momentum tensor and the baryon current
carried by the fluid part, respectively.
%% The source terms $J^\nu$ and $\rho$
%% are the transfer of energy--momentum and baryon number
%% from the JAM particles to the fluids.
For the source terms we consider
the absorption of particles into the fluids:
\begin{align}
  J^\mu(t,\bm{r}) &= \frac1{\Delta t} \sum_i p_i^\mu G(\bm{r} - \bm{r}_i), &
  \rho(t,\bm{r}) &= \frac1{\Delta t} \sum_i B_i G(\bm{r} - \bm{r}_i),
\end{align}
where $p_i^\mu$, $\bm{r}_i$ and $B_i$ denote the momentum, position and baryon number of the $i$-th particle,
and the sum runs over the absorbed particles which decay within the time step $\Delta t$
in the region where the energy density $e$ satisfies $e > e_\mathrm{f}$.
%% The absorbed particles are removed from JAM cascade.
The smearing profile $G(\bm{r})$ is given by the Lorentz contracted Gaussian profile
%% with the standard deviation in the rest frame being $\sigma = 0.5\ \text{fm}$.
with the width $\sigma = 0.5\ \text{fm}$.
In a corona region where $e<e_\mathrm{p}$,
fluids are converted to JAM particles using the positive contribution of the Cooper--Frye formula:
%% fluids are converted to JAM particles using the Monte--Carlo sampling
%% of the Poisson distribution whose mean particle number
%% is given by the positive contribution of the Cooper--Frye formula:
\begin{align}
  \Delta N_i &= \frac{g_i}{(2\pi)^3} \int \frac{d^3p}E \frac{(\Delta \sigma\cdot p) \theta(\Delta \sigma\cdot p)}{\exp[(p\cdot u - \mu_i)/T] \pm 1},
\end{align}
where $\Delta\sigma_\mu$ is the surface element of the particlization hypersurface $e=e_\mathrm{p}$,
the coefficients $g_i$ and $\mu_i$ are the spin degeneracy and the chemical potential
of the $i$-th hadron species, respectively,
and $\pm$ is $+$ for fermions and $-$ for bosons.
%% The energy--momentum and baryon number of the converted fluid elements
%% are then initialized to zero.
In this way, we simultaneously solve the JAM cascade and ideal hydrodynamics which are
dynamically coupled to each other through the source terms in the core, $e>e_\mathrm{f}$,
and the Cooper--Frye formula in the corona, $e<e_\mathrm{p}$.

%%%%%%%%%%%%%%%%%%%%%%%%%%%%%%%%%%%%%%%%%%%%%%%%%%%%%%%%%%%%%%%%%%%%%%%%%%%%%%

\section{Results}
\begin{figure}[htb]%
  \centering\vspace{-2mm}%
  \includegraphics[width=0.99\textwidth]{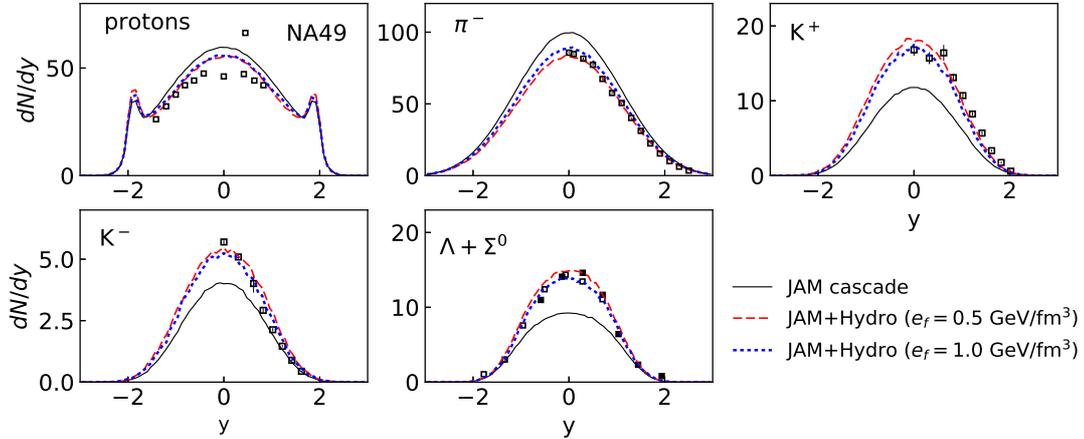}%
  \vspace{-2mm}%
  \caption{
    The rapidity distributions of identified hadrons are plotted
    for central Pb+Pb collisions at $\sqrt{s_\mathrm{NN}} = 6.4\ \text{GeV}$
    for the fluidization energy $e_\mathrm{f} = 0.5$ (red dashed) and $1.0\ \text{GeV/fm$^3$}$ (blue dotted).
    The black solid lines and points are from JAM cascade and experimental data~\cite{ref:rapidity-data}, respectively.}
  \label{fig:rapidity}
\end{figure}
\begin{figure}[htb]%
  \centering\vspace{-2mm}%
  \includegraphics[width=0.99\textwidth]{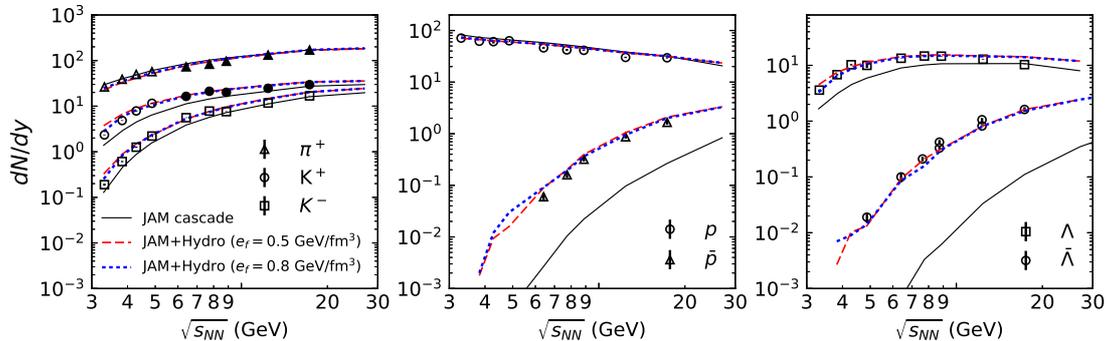}%
  \vspace{-2mm}%
  \caption{The beam-energy dependence of identified hadron yields are plotted
  for the fluidization energy $e_\mathrm{f} = 0.5$ (red dashed line)
  and $0.8\ \text{GeV/fm$^3$}$ (blue dotted line).
  The JAM cascade results and experimental data~\cite{ref:yield-data} are
  shown by the black solid line and points, respectively.}
  \label{fig:yields}
\end{figure}
\begin{figure}[htb]
  \centering
  \vspace{-3mm}
  \includegraphics[width=0.5\textwidth]{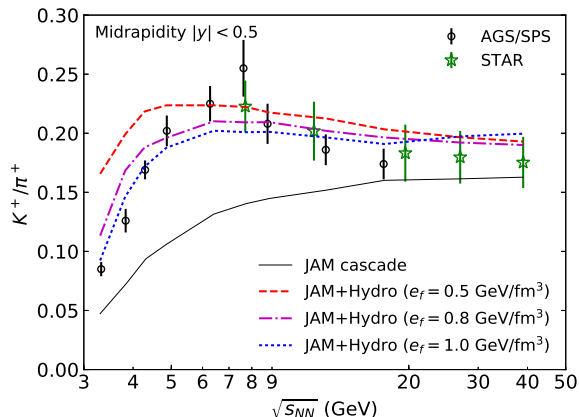}
  \vspace{-2mm}
  \caption{The particle ratio for positively charged kaons and pions
    are plotted as a function of the beam energy
    for each fluidization energy $e_\mathrm{f} = 0.5$ (red dashed line),
    $0.8$ (purple chain line) and $1.0$ (blue dotted line) $\text{GeV}/\text{fm}^3$.
    The black solid line and points are from the JAM cascade and experiments~\cite{ref:ratio-data}, respectively.}
  \label{fig:kpiratio}
\end{figure}
Using the JAM+Hydro model, we performed calculations of
Au+Au collisions ($\sqrt{s_\mathrm{NN}} = 2.7$--$4.9\ \text{GeV}$)
and Pb+Pb collisions ($\sqrt{s_\mathrm{NN}} = 6.4$--$17.3\ \text{GeV}$)
with the impact parameter $b<4.0\ \text{fm}$ for 0--7\% centrality at $\sqrt{s_\mathrm{NN}} = 6.4$--$12.4\ \text{GeV}$
or otherwise $b<3.4\ \text{fm}$ for 0--5\% centrality.
Figure~\ref{fig:rapidity} shows rapidity distributions of identified hadrons.
Compared to the JAM cascade model, the rapidity distributions
is improved by the JAM+Hydro model for all the hadron species.
The change of the protons from the JAM cascade model to the JAM+Hydro model
is small for both fluidization energy $e_\mathrm{f}=0.5$ and $1.0\ \text{GeV}/\text{fm}^3$,
which means that the stopping power of the two nuclei is unaffected by the introduction of hydrodynamic description.
We also notice that the negatively charged pions are suppressed while the strange hadrons are enhanced
by the hydrodynamic evolution because the yields become closer to the equilibrium values
by thermalization forced by the conversion of particles to fluids.
The beam-energy dependence of identified hadron yields is shown in Fig.~\ref{fig:yields}.
The JAM cascade slightly overestimates pions
and underestimates strange particles.
This situation is resolved by the JAM+hydro model,
and good agreement with the data is obtained.
Figure~\ref{fig:kpiratio} shows
the $K^+/\pi^+$ ratio as a function of the beam energy.
The JAM+Hydro prediction is significantly
improved from the JAM cascade prediction.
The ratio is sensitive to the fluidization energy $e_\mathrm{f}$ at AGS energies,
and larger values of $e_\mathrm{f}$ improve the description.
The ratio at higher beam energies is not so much affected by $e_\mathrm{f}$.
Also the transverse mass spectra of identified hadrons
are in good agreement with experimental data~\cite{ref:PAPER}.

%%%%%%%%%%%%%%%%%%%%%%%%%%%%%%%%%%%%%%%%%%%%%%%%%%%%%%%%%%%%%%%%%%%%%%%%%%%%%%

\section{Summary}
For high-energy nuclear collisions in a wide range of the beam energy
to explore the high baryon density region of the QCD phase diagram,
we developed a new dynamical model, JAM+Hydro,
in which a hadronic transport model JAM and ideal hydrodynamics
are dynamically integrated by putting emphasis on
dynamical initialization,
dynamical core--corona separation
and dynamical coupling of the models.
We performed calculations of central Au+Au/Pb+Pb collisions
in the beam-energy range of $\sqrt{s_\mathrm{NN}} = 2.7$--$17.3\ \text{GeV}$
and obtained significantly improved results of the beam-energy dependence
of identified hadron yields and the particle ratio of positively charged kaons and pions,
which can explain the experimental data well.


\begin{thebibliography}{9}
\bibitem{ref:critical-point}
  %% M.~Asakawa and K.~Yazaki, Nucl.\ Phys.\ A {\bf 504}, 668 (1989);
  %% D.~H.~Rischke, Prog.\ Part.\ Nucl.\ Phys.\ {\bf 52}, 197 (2004);
  %% M.~A.~Stephanov, Prog.\ Theor.\ Phys.\ Suppl.\ {\bf 153}, 139 (2004) [Int.\ J.\ Mod.\ Phys.\ A {\bf 20}, 4387 (2005)];
  %% K.~Fukushima and C.~Sasaki, Prog.\ Part.\ Nucl.\ Phys.\ {\bf 72}, 99 (2013).
  M.~Asakawa and K.~Yazaki, Nucl.\ Phys.\ A {\bf 504}, 668 (1989);
  D.~H.~Rischke, Prog.\ Part.\ Nucl.\ Phys.\ {\bf 52}, 197 (2004);
  M.~A.~Stephanov, Prog.\ Theor.\ Phys.\ Suppl.\ {\bf 153}, 139 (2004).
\bibitem{ref:JAM}
  Y.~Nara, N.~Otuka, A.~Ohnishi, K.~Niita, and S.~Chiba, Phys.\ Rev.\ C {\bf 61}, 024901 (2000).
\bibitem{ref:PAPER}
  Y.~Akamatsu {\it et al.}, Phys.\ Rev.\ C {\bf 98}, no. 2, 024909 (2018).
\bibitem{ref:dynamical-init}
  %% M.~Okai, K.~Kawaguchi, Y.~Tachibana, and T.~Hirano, Phys.\ Rev.\ C {\bf 95}, no. 5, 054914 (2017);
  %% C.~Shen and B.~Schenke, Phys.\ Rev.\ C {\bf 97}, no. 2, 024907 (2018).
  M.~Okai {\it et al.}, Phys.\ Rev.\ C {\bf 95}, 054914 (2017);
  C.~Shen {\it et al.}, Phys.\ Rev.\ C {\bf 97}, 024907 (2018).
\bibitem{ref:core-corona-sep}
  K.~Werner, Phys.\ Rev.\ Lett. {\bf 98}, 152301 (2007).
%% \bibitem{format} (Ref format) F. Author, S. Author, and T. Author, Abbreviated
%%   journal title \textbf{volume in bold face}, initial page or article
%%   number (year of publication).
%% \bibitem{instructions} More abbreviations of journal titles are listed
%%   in ``Instructions for Preparation of Manuscript", which is available
%%   at our Web site (http://jpsj.ipap.jp).
\bibitem{ref:EOSQ}
  %%  J. Sollfrank, P. Huovinen, M. Kataja, P. V. Ruuskanen,
  %% M. Prakash, and R. Venugopalan, Phys. Rev. C 55, 392 (1997);
  %% P. F. Kolb, J. Sollfrank, and U. W. Heinz, Phys. Lett. B 459, 667
  %% (1999); P. F. Kolb, J. Sollfrank, and U. W. Heinz, Phys. Rev. C
  %% 62, 054909 (2000).
  J.~Sollfrank {\it et al.}, Phys.\ Rev.\ C {\bf 55}, 392 (1997).
\bibitem{ref:rapidity-data}
  C.~Alt {\it et al.} [NA49 Collaboration], Phys.\ Rev.\ C {\bf 77}, 024903 (2008).
  C.~Blume [NA49 Collaboration], J.\ Phys.\ G {\bf 34}, S951 (2007).
  C.~Alt {\it et al.} [NA49 Collaboration], Phys.\ Rev.\ C {\bf 78}, 034918 (2008).
  %% NA49 Collaboration, Phys.\ Rev.\ C {\bf 77}, 024903 (2008);
  %% J.\ Phys.\ G {\bf 34}, S951 (2007);
  %% Phys.\ Rev.\ C {\bf 78}, 034918 (2008).
\bibitem{ref:yield-data}
  L.~Ahle {\it et al.} [E866 and E917 Collaborations], Phys.\ Lett.\ B {\bf 476}, 1 (2000);
  %C. Blume, M. Gazdzicki, B. Lungwitz, M. Mitrovski, P. Seyboth, and H. Stroebele, NA49 Compilation, https://edms.cern.ch/document/1075059;
  C.~Blume {\it et al.}, https://edms.cern.ch/document/1075059;
  C.~Blume and C. Markert, Prog.\ Part.\ Nucl.\ Phys.\ {\bf 66}, 834 (2011);
  L.~Ahle {\it et al.} [E-802 Collaboration], Phys.\ Rev.\ C {\bf 57}, no. 2, R466 (1998);
  J.~L.~Klay {\it et al.} [E895 Collaboration], Phys.\ Rev.\ Lett.\ {\bf 88}, 102301 (2002).
\bibitem{ref:ratio-data}
  %% L. Adamczyk {\it et al.} [STAR Collaboration], Phys.\ Rev.\ C {\bf 96}, no. 4, 044904 (2017);
  %% A. Andronic, P. Braun-Munzinger, and J. Stachel, Nucl. Phys. A 772, 167 (2006).
  %% A. Andronic, P. Braun-Munzinger, and J. Stachel, Phys. Lett. B 673, 142 (2009) Erratum: [Phys. Lett. B 678, 516 (2009)].
  L. Adamczyk {\it et al.} [STAR Collaboration], Phys.\ Rev.\ C {\bf 96}, 044904 (2017);
  A. Andronic {\it et al.}, Nucl.\ Phys.\ A {\bf 772}, 167 (2006);
  A. Andronic {\it et al.}, Phys.\ Lett.\ B {\bf 673}, 142 (2009) [Phys.\ Lett.\ B {\bf 678}, 516 (2009)].
\end{thebibliography}
\end{document}